\RequirePackage{fix-cm}
\documentclass{amsart}                     
\usepackage{graphicx}
\usepackage[utf8]{inputenc}
\usepackage[T1]{fontenc}
\usepackage{float}
\usepackage{subfigure}

\newcommand{\f}{\frac}

\def\re#1{(\ref{#1})}  
\begin{document}

\title{Thermodynamical consistency of the Dual Phase Lag heat conduction equation
}


\author{Róbert Kovács$^{123}$         \and
        Péter Ván$^{123}$
}



\address{$^1$Department of Theoretical Physics, Wigner Research Centre for Physics,
Institute for Particle and Nuclear Physics, Budapest, Hungary and
$^2$Department of Energy Engineering, Faculty of Mechanical Engineering, BME, Budapest, Hungary and
$^3$Montavid Thermodynamic Research Group}


\date{Received: date / Accepted: date}

\maketitle

\begin{abstract}
Dual phase lag equation for heat conduction is analyzed from the point of view of non-equilibrium thermodynamics. Its first order Taylor series expansion is consistent with the second law as long as the two relaxation times are not negative. 


\flushleft{
The work was supported by the grants National Research, Development and Innovation Office: NKFIH 116197 and 124366.}

\end{abstract}

\section{Introduction}
\label{intro}

Deviation from Fourier's law was observed in low temperature experiments \cite{Pesh44,Rog71a,JacWalMcN70,FairLaWill47,Acketal66}. These  experiments were performed with uniform homogeneous test samples where phonon based kinetic theory is a suitable theoretical background \cite{Max1867,GuyKru66a1,GK66,DreStr93a}. However, there are efforts to find simpler and more universal explanations \cite{Cattaneo58,Vernotte58,JosPre89,JosPre90a,JouCimm16,SelEta16b}. In heterogeneous materials at room temperature the experiments did not confirm wave like propagation \cite{HerBec00b,RoeEta03a}, but a different form of generalized heat conduction was confirmed, given by the following equation in one spatial dimension:
\begin{equation}
\tau \partial_{tt} T + \partial_t T = \alpha \partial_{xx} T + \kappa^2 \partial_{txx} T,
\label{gk_temp}
\end{equation}
Here $T$ is the temperature, $\alpha$ is the thermal diffusivity coefficient, $\tau$ is the relaxation time and $\kappa$ is an additional dissipation coefficient. If $\kappa=0$, the equation above is usually interpreted as a telegraph equation where damped wave propagation is observed. The dissipation coefficient $\kappa$ can be interpreted as a length parameter, e.g. in rarefied gases it is proportional to the phonon mean free path $l$.

There are two different constitutive relations from where this partial differential equation can be derived. The first order differential Dual Phase Lag constitutive equation (DPL) has the same form as the so called Jeffreys type (JT) constitutive equation, except the different interpretation of the coefficients: 
\begin{equation}
\tau \partial_t q+q=-\lambda \partial_x T - \hat \tau \lambda \partial_{xt} T.
\label{jeffequat}
\end{equation}

This is the DPL form and $\hat \tau$ is the temperature relaxation time. With the definition of the effective thermal conductivity  $\hat \lambda$, as $\hat \lambda \tau = \lambda\hat{\tau}$ it may be rewritten into the JT form. This last form was introduced by Joseph and Preziosi by an analogous equation in rheology \cite{Verhas97} assuming a memory effect between the temperature gradient and the related heat flux in an integral form  \cite{JosPre89}. The DPL form was introduced by Tzou assuming a memory effect by a direct time delay in the Fourier's law \cite{Tzou95,Tzou96}. The interpretation and the role of the second law is different in the two cases.

The second constitutive equation, the Guyer-Krumhansl equation, was derived by the linearization of the Boltzmann equation with the double relaxation time Callaway collision integral. It has the following form: 
\begin{equation}
\tau \partial_t q+q=-\lambda \partial_x T + \kappa^2 \partial_{xx} q.
\label{gkeq}
\end{equation}

In their original derivation none of the above mentioned equations were analyzed considering their admissibility to the second law. In case of the Jeffreys type equation the delay was introduced to the Fourier's law by memory integrals, therefore it was plausible to assume a kind of weak compatibility due to the idea of fading memory as it has been investigated later \cite{Fabetal14}. In case of the dual phase lag concept, with delay equations, there is no way to check this important requirement. For the Guyer-Krumhansl equation it is possible to investigate the compatibility directly, because the momentum series expansion of the kinetic theory and also the related phenomenological theory, namely Rational Extended Thermodynamics has a proper form of the second law inequality. However, in that theory the GK form is not conceptually important, because the compatibility is required only for the basic hyperbolic system \cite{MulRug98,DreStr93a}.  

In this paper, first we critically survey the dual phase lag concept. Then we shortly show a simple derivation where both Guyer-Krumhansl equation and Jeffreys type equations can be obtained from uniform thermodynamic principles extending the local equilibrium. In that framework the second law and the other thermodynamic requirements can be considered. Finally we compare the equations and analyze the thermodynamic admissibility of the dual phase lag.

\section{Dual phase lag concept}
\label{DPLC}

Different time delay concepts to obtain memory effects of heat conduction appear in the literature independently in several cases. E.g. in \cite{Taitel72}, Taitel used a single phase lag to obtain the MCV equation. The DPL equation was proposed by Tzou \cite{Tzou95,Tzou96} in order to account time delay effect and to generalize the single phase lag concept leading to Maxwell-Cattaneo-Vernotte (MCV) equation \cite{Max1867,Cattaneo58,Vernotte58}. It is formulated as follows. Let us suppose a time lag $\tau_q$ for the heat flux and a different one for the temperature gradient, denoted by $\tau_T$, the constitutive equation in one dimension  reads as
\begin{equation}
q(x,t+\tau_q) = -\lambda \partial_x T (x, t + \tau _T),
\label{eq:dpl}
\end{equation}
where $\tau_q>0$ and $\tau_T>0$ relaxation times are strictly positive. The DPL equation in 1+1 D is derived by Taylor series expansion:
\begin{equation}
q (x,t) + \tau_q \partial_t q(x,t) \approx -\lambda \left ( \partial_x T (x,t) + \tau_T \partial_{tx} T (x,t) \right ).
\label{eq:dpl_type1}
\end{equation}
Here, considering $\tau_T=0$ results in  the MCV equation. In other versions of DPL, namely the "type 2 DPL" (DPL2) and "type 3 DPL" (DPL3) the Taylor series expansion is continued. In case of DPL2, the constitutive equation according to \cite{XuSeffen08,SahooEtal14} reads as
\begin{equation}
q (x,t) + \tau_q \partial_t q(x,t) = -\lambda \left ( \partial_x T (x,t) + \tau_T \partial_{tx} T  (x,t) + \f{\tau_T^2}{2} \partial_{ttx} T (x,t)\right ).
\end{equation}
The DPL3 model is extended by the second order time derivative of the heat flux:
\begin{equation}
q (x,t) + \tau_q \partial_t q(x,t) +\f{\tau_q^2}{2} \partial_{tt} q = -\lambda \left ( \partial_x T (x,t) + \tau_T \partial_{tx} T  (x,t) + \f{\tau_T^2}{2} \partial_{ttx} T (x,t)\right ).
\end{equation}

The DPL equation is very simple, the intuitive idea is sound and the first approximation become a popular model especially on the field of biological heat conduction, see e.g. \cite{Ant05meat,AfrinEtal12,Zhang09,ChenZan09,AfrZhang11,LiuChen10}. Based on a fractal approach, the idea behind the dual phase lag concept is further extended by Ezzat et al. to a three phase lag approach \cite{EzEtal13} and applied by Akbarzadeh and Pasini along with the DPL model \cite{AkbPas14}. These generalizations show how easy to apply these ideas. However, the original delay type constitutive equation is less applicable, important conceptual and numerical problems arise. In order to overcome these difficulties the second law compatibility was suggested as a possible way of improvement \cite{Fabetal14,FabLaz14a,FabEtal16}. In the next section we introduce a framework where this important conceptual question can be considered beyond local equilibrium.

\section{Derivation based on non-equilibrium thermodynamics}
\label{DNET}

In this section only the essential steps of the derivation are mention briefly and only in 1+1D, a detailed, more general  description is published in \cite{Van01a,VanFul12,KovVan15}.  

We consider rigid heat conductors and our starting point is the conservation of internal energy:
\begin{equation}
\rho \partial_t e + \partial^i q^i =0,
\label{inten_bal}
\end{equation}
where $\partial_t$ is the partial time derivative, $q^i$ is the heat flux, $e$ denotes the specific internal energy and it can be expressed as $e=cT$ where $c$ is the  specific heat. Einstein summation convention is applied. The second law of thermodynamics is exploited in this framework, therefore the balance of entropy inequality is used:
\begin{equation}
\sigma_s = \rho \partial_t s + \partial^i J^i \geq 0,
\label{s_bal}
\end{equation}
where $s$ denotes the specific entropy, $J^i$ is the entropy flux and $\sigma_s$ is the entropy production. Thermodynamic stability is preserved by assuming a quadratic deviation from the local equilibrium in the entropy density \cite{Verhas97},
\begin{equation}
s(e, \xi) = s_{{eq}}(e) - \frac{m}{2} \xi^2,
\label{neqs}
\end{equation}
where  $m$ is a positive scalar material parameter, $\xi^i$ is a vectorial internal variable. In case of heat conduction it is customary to choose heat flux,  $q^i$, as internal variable \cite{MulRug98,SelEta16b}, i.e. $\xi^i = q^i$. This way one can accommodate to the kinetic theory derivation of continuum equations. Here we do not use this simplification.

Considering that densities and fluxes are parts of a space-time quantity also in nonrelativistic spacetime, is is natural to assume that the classical entropy current is generalized considering the simplest deviation from the usual form, with the help of Ny\'iri multipliers \cite{Nyiri89}. 
\begin{equation}
J^i = b^{ij} q^j,
\label{neqJ}
\end{equation}
where $b^{ij}$ is a $2^{{nd}}$ order tensor called current multiplier \cite{Nyiri91,KovVan15}. For the classical entropy flux $b^{ij}=\f{1}{T}\delta^{ij}$. Then the entropy production $\sigma_s$ is calculated as follows:
\begin{eqnarray}
\sigma_s = \rho \partial_t s + \partial^i J^i &=& \nonumber \\
 \rho \left ( \partial_e s \partial_t e +\partial_{\xi^i} s \partial_t \xi^i \right ) + \partial^i \left ( b^{ij} \xi^j \right ) &=& \nonumber \\
 \rho \left ( \f{1}{T} \partial_t e - m \xi^i \partial_t \xi^i \right ) + (\partial^i b^{ij})\xi^j + b^{ij}(\partial^i \xi^j) &=& \nonumber \\
 \xi^j  \partial^i b^{ij} - \rho m q^j\partial_t q^j   + \partial^j q^i \left ( b^{ij} - \f{1}{T}\delta^{ij} \right ) \geq 0.
\label{entrpr}
\end{eqnarray}
Here $\delta^{ij}$ stands for the Kronecker symbol. Let us consider an isotropic material and  one dimensional propagation. Then only the vectorial thermodynamic fluxes and forces are related and the linear solution of inequality (\ref{entrpr}) is
\begin{eqnarray}
m \partial_t \xi &=& -l_1 \xi + l_{12} \partial_x b, \label{ce1}\\
q &=& -l_{21} \xi + l_2 \partial_x b, \label{ce2}\\
b- \frac{1}{T} &=&  l_{3} \partial_x q. \label{ce3}
\end{eqnarray}

According to the second law, the following conditions are required for the coefficients $l_1, l_1, l_{12}, l_{21}, l_3$:
\begin{equation} 
l_1\geq 0, \quad
l_2\geq 0, \quad 
l_3\geq 0, \quad  \quad
L=l_1l_2-l_{12}l_{21} \geq 0.
\label{coeffreq}
\end{equation}
Eliminating the current multiplier $b$ and the internal variable $\xi$ by substitution, the general heat conduction equation for $q$ is obtained.
\begin{equation}
m \partial_t q + l_1 q  - L l_3 \partial_{xx} q - m l_2 l_3 \partial_{xxt} q  = L \partial_x \frac{1}{T}+m l_2 \partial_{xt} \f{1}{T} .
\label{bigone}
\end{equation}

Eq. (\ref{bigone}) consists of the Fourier and MCV equations as particular cases. In our case, the following two equations have to be emphasized. 
\begin{enumerate}
\item The Guyer--Krumhansl equation (GK)  can be obtained when $l_2=0$:
\begin{equation} 
m \partial_tq+l_1q-L l_3\partial_{xx}q=L \partial_x \frac{1}{T}, 
\label{GKt}\end{equation}
that is
\begin{equation}
\tau_q \partial_t q+q=-\lambda\partial_x T + \kappa^2 \partial_{xx}q,
\label{GKlin}\end{equation}
where the coefficients are identified as $\tau_q=\frac{m}{l_1}$, $\lambda=\frac{L}{l_1 T^2}$, $\kappa^2=\frac{L l_3}{l_1}$. All of them are positive.

\item Jeffreys type or Dual Phase Lag thermodynamical equation appears if $l_3=0$:
\begin{equation} 
m \partial_tq+l_1 q = L\partial_x \frac{1}{T} +  m l_2 \partial_{xt}\frac{1}{T}.
\label{JTt}\end{equation}
This nonlinear equation is called thermodynamical, because we have applied the natural thermodynamic force for thermal interaction, the gradient of the reciprocal temperature, instead of the usual temperature gradient. This appears in the last two terms. A linearization around a reference temperature leads to 
\begin{equation}
\tau_q \partial_t q+q=-\lambda \partial_x T +  \chi^2 \partial_{xt}T,
\label{JTlin}
\end{equation}
where the coefficients are related to thermodynamic origins as previously, except the last one, $\chi^2 = \f{ml_2}{l_1}$. It is clear, that $\chi^2  >0$  because of (\ref{coeffreq}). Furthermore we can decompose this coefficient according to the DPL and JT interpretations introducing $\tau_T = \f{ml_2}{L}$ in case of the DPL interpretation and $\lambda_1 =  l_2$ for the JT one as  $\chi^2  = \lambda \tau_T= \lambda_1\tau_q$.
\end{enumerate}

\section{Criticism of DPL}
\label{CRIT}

The DPL and also the Guyer-Krumhansl equations are criticized from several different points of view. 

{\it Neqative temperatures.} Rukolaine \cite{Ruk14,Ruk17} obtained an analytical solution for DPL equation assuming a Gaussian initial condition. According to these calculations, the solutions present an unphysical behavior of temperature history, it goes into the negative domain. Zhukovsky \cite{Zhu16a,Zhu16b} achieved a similar conclusion for the GK equation. Wang et al. \cite{Wang11nonFou} tested the thermomass and DPL among other different heat conduction models by calculating Taitel's problem \cite{Taitel72}. An inconsistent, unphysical behavior is shown, the temperature achieves the negative domain again. This is actually an old problem, also mentioned several times for the Maxwell-Cattaneo-Vernotte equation. In fact it is not paradoxical, when we are considering relative temperatures as in the linearized \re{GKlin} and \re{JTlin} equations, because the wave-like propagation mode. However, with the real thermodynamic models with the gradient of the absolute reciprocal temperature as thermodynamic force, like  \re{GKt} and \re{JTt}, this should be investigated.

{\it Time shift paradox.} The delayed  form of the DPL equation, \re{eq:dpl}, directly contradicts to the requirement of the homogeneity of time. One can transform the equation to a single phase lag equation by shifting the zero point of $t$, the starting point of the  time measurement. The proper mathematical representation of nonrelativistic time considering all expected properties is a one dimensional affine space and with that model \re{eq:dpl} cannot have two relaxation times, only their difference  is that plays a role.

{\it Mathematical question.} In the literature, the DPL-type constitutive equations are analyzed in order to prove the uniqueness and well-posedness of a process driven by such constitutive equation, too. It is found by Fabrizio et al. \cite{Fabetal14,FabLaz14a,FabEtal16} that there are mathematical conditions beyond the physical ones to obtain an exponentially stable equilibrium solution for DPL equation. Such condition requires negative  time delay (called as retarded effect) between the heat flux and temperature gradient, i.e. $\tau_q - \tau_T \leq 0$ which excludes the opposite case. It is important because Tzou in \cite{Tzou96} directly interprets both case with the cause -- effect concept, i.e. the quantity with the higher relaxation time is the effect caused by the other one. As Fabrizio states \cite{Fabetal14}, the DPL model can be rewritten with the time delay difference $\tau_d := \tau_q - \tau_T$ which leads to a single phase lag model but here the temperature gradient have a relaxation time, with $\tau_d <0$. The opposite case, i.e. $\tau_d >0$, is mathematically ill-posed which enlightens the validity of MCV equation but excludes equations based on arbitrary Taylor series expansion. Quintanilla et al. \cite{Quin07,Dreetal09} obtains the same conclusion regarding the relaxation times and the ill-posedness.

However, exponential stability seems to be too strict requirement and the asymptotic stability of homogeneous equilibrium  is satisfied only with nonnegative coefficients in equations \re{inten_bal}, \re{GKlin} and \re{JTlin}, whenever the sign restrictions from the entropy inequality, \re{coeffreq}, and thermodynamic stability, $c>0$, is satisfied, see e.g. \cite{Fabetal14}. The requirement of asymptotic stability is also reasonable from a physical  point of view \cite{Van95a,Mato04b,VanBir08a,Van09a}

{\it Second law.} Fabrizio and coworkes also checked thermodynamical restriction. It was tested by using Clausius-Duhem inequality on cyclic histories. The conditions of exponential stability for the DPL equation  turned out to be too strict. It requires $\tau_T -\tau_q\geq 0$ and the second law alone does not warrant it \cite{Fabetal14,FabLaz14a}. On the other hand the experimental evidences mostly show that this difference is negative.  In the work of Liu and Chen \cite{LiuChen10}, the DPL equation is fitted to experimental results with $\tau_T>\tau_q$ in every case and in heterogeneous materials there is a similar  situation \cite{PMMar17conf,Botetal16,Vanetal17}. It draws attention to the practical aspects which emphasized further in the next section.


\section{Experiment revisited}
\label{ExpR}

In this section the experiment of Tang et al. is discussed in detail \cite{TanEtal07}. The forthcoming measurement results are evaluated with the help of the Fourier and DPL equations by Tang et al. Instead of the DPL model, we reevaluated the results using the Guyer-Krumhansl equation.

\begin{figure}[H]
\centering
\includegraphics[width=11cm,height=6cm]{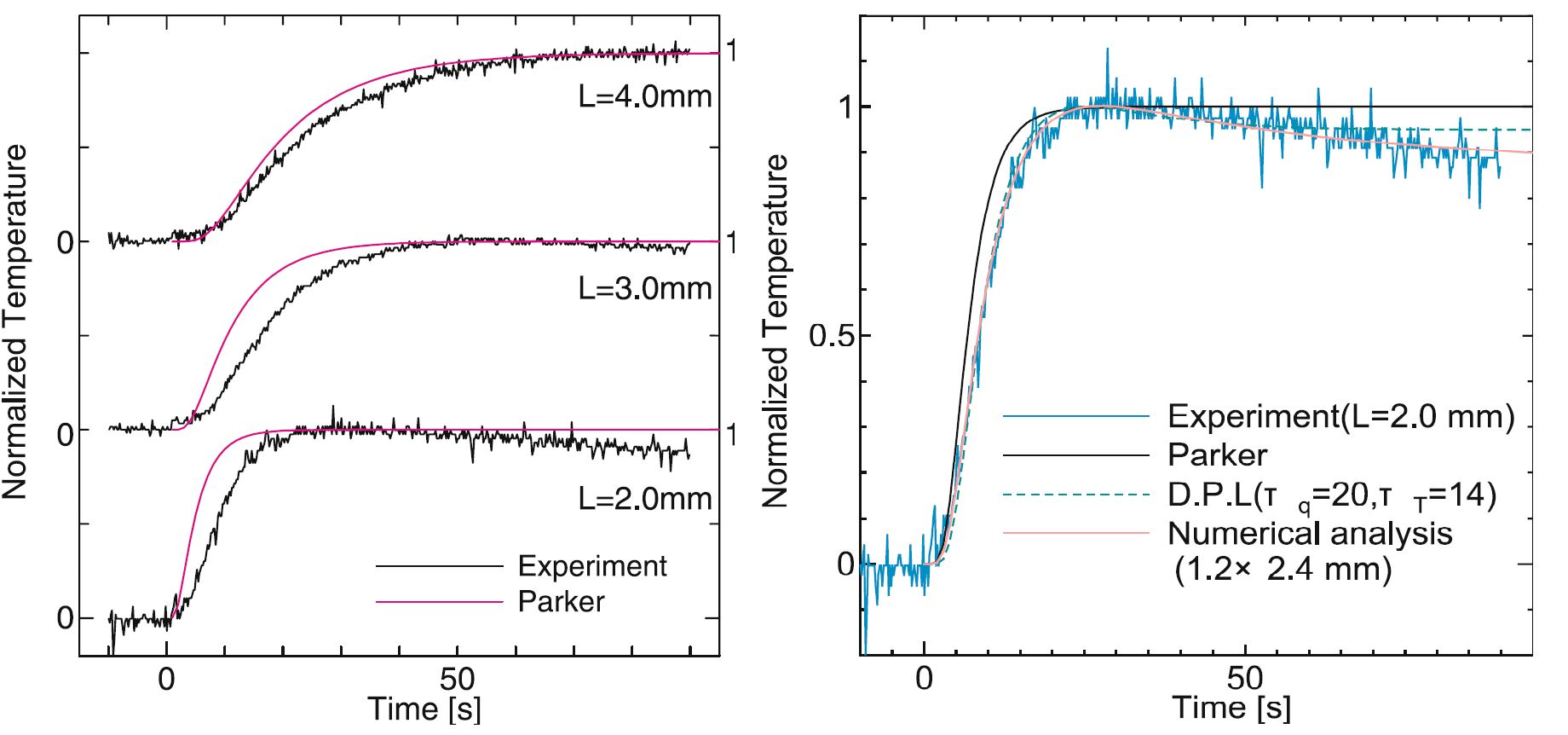}
\caption{The original figure about the experimental results, Fourier and DPL fitting from Tang et al. \cite{TanEtal07}. The name of Parker refers to the classical theory, based on Fourier's law.}
\label{fig:bio8}
\end{figure}

It was also a heat pulse experiment but the sample is a processed meat similar to the one from the famous paper of Mitra et al. \cite{MitEta95}. The diameter of the samples is $10 \ mm$, their shape is cylindrical. The experiment is repeated with three different sample lengths, i.e. $L = 2$, $3$ and $4 \ mm$. The heat pulse lasts $1 s$. The front end of each specimen is blackened by using graphite spray. At the rear end, Cu film is applied with thickness of $0.01 \ mm$. The results can be seen on Fig. \ref{fig:bio8}, \cite{TanEtal07}.

At first sight, the non-Fourier behavior seems to be clear from the deviation between the fitting and measured data (Fig. \ref{fig:bio8}). The cooling of the specimen is relevant in the case where the length $L$ is $2$ and $3 \ mm$. The evaluation of such measured result is the same as in case of our experiments \cite{Botetal16,Vanetal17} due to the same arrangement. However, in these cases the dimensionless temperature cannot reach one, not even theoretically, due to cooling. Thus in the second and third cases the fitting should be reconsidered.

\begin{figure}[H]
     \begin{center}
        \subfigure[Fourier fit for the first case; $\alpha = 1.1 \cdot 10^{-7} \f{m^2}{s}$]{%
            \label{fig:tan1}
            \includegraphics[width=6cm,height=3.5cm]{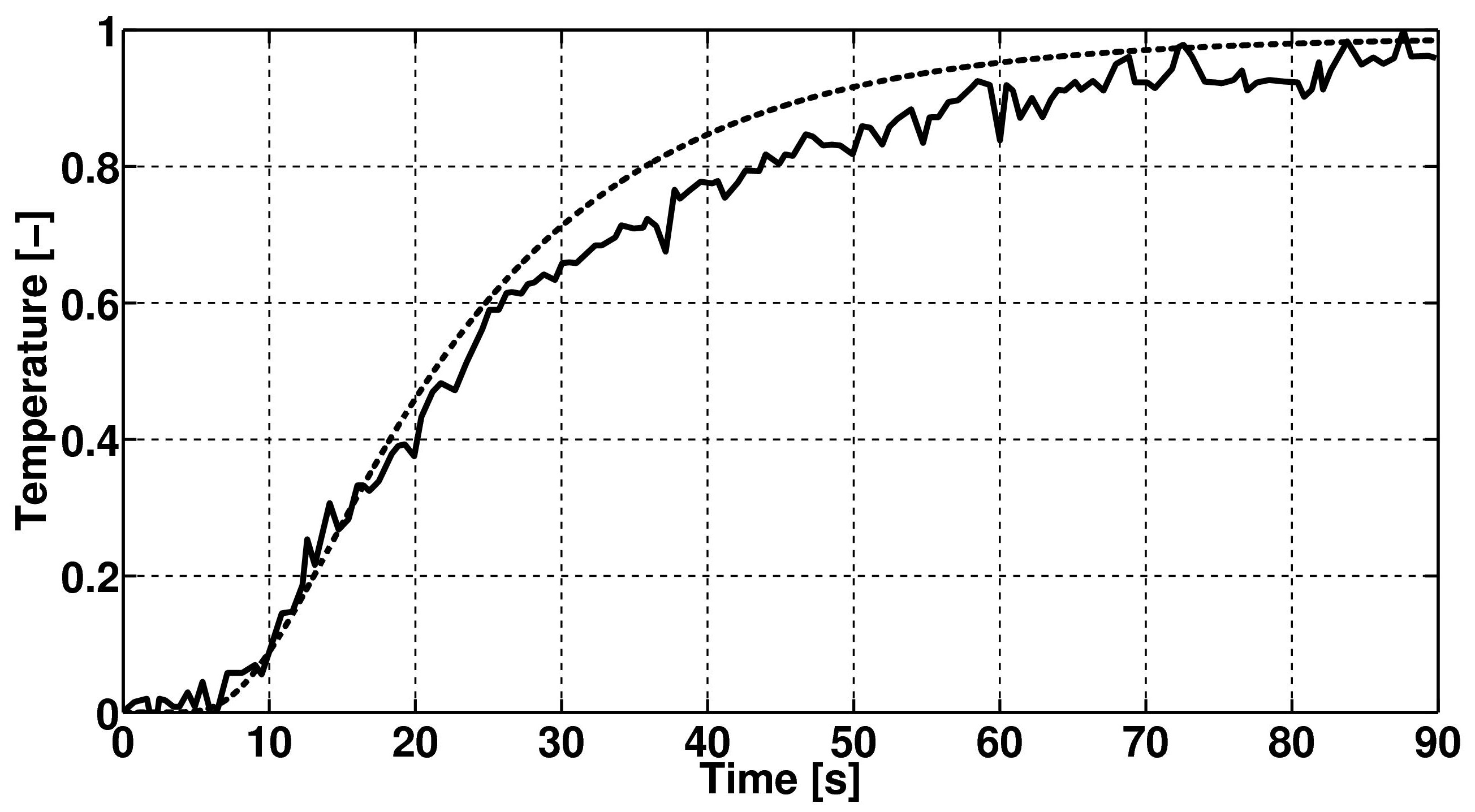}
        }%
        \subfigure[GK fit for the first case $\alpha = 9.4 \cdot 10^{-8} \f{m^2}{s}$, $\tau_q = 3.574 s$, $\kappa^2 = 5.44 \cdot 10^{-7} m^2$]{%
           \label{fig:tan2}
           \includegraphics[width=6cm,height=3.5cm]{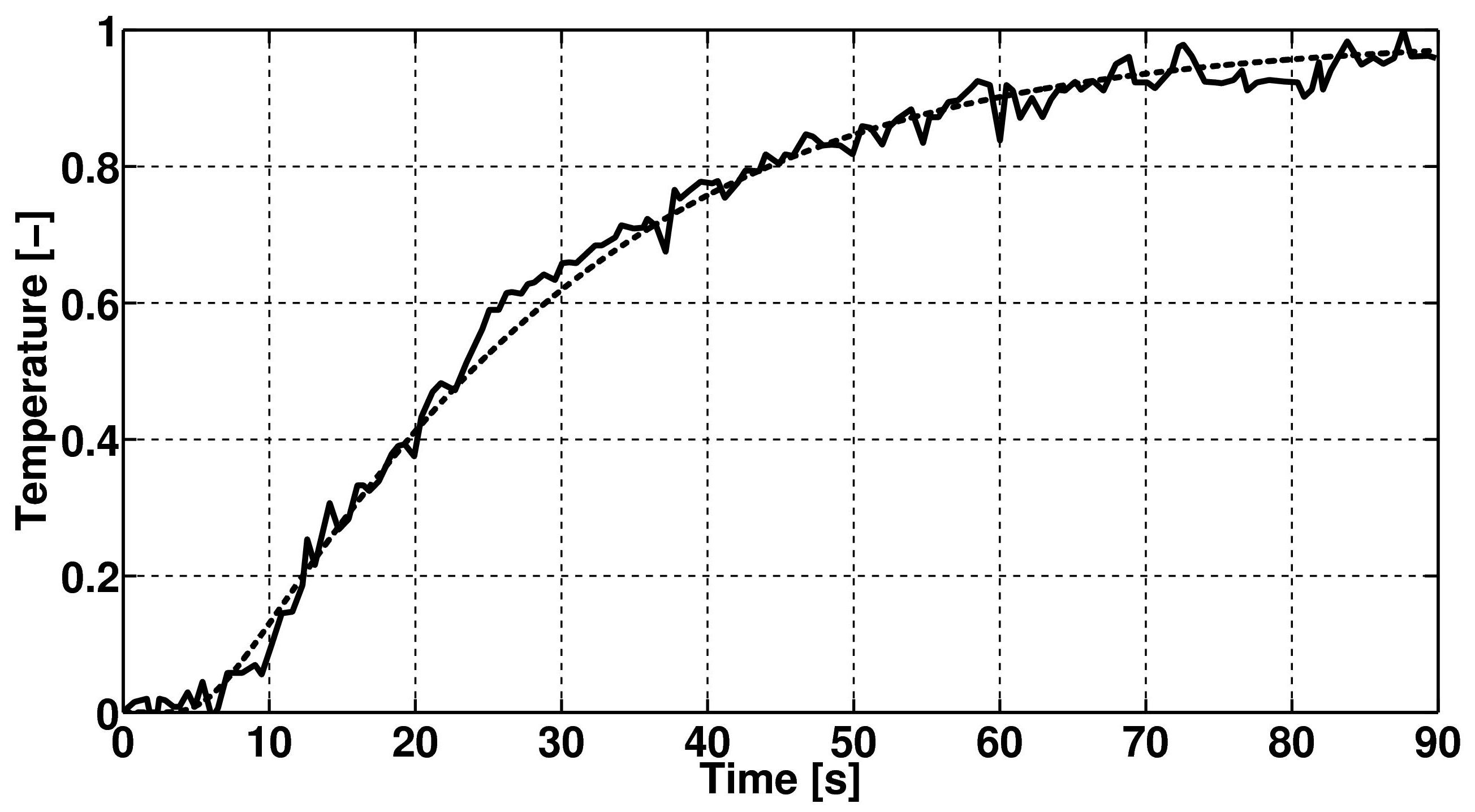}
        } \\
	\subfigure[Fourier fit for the second case; $\alpha = 7.7 \cdot 10^{-8} \f{m^2}{s}$]{%
            \label{fig:tan3}
            \includegraphics[width=6cm,height=3.5cm]{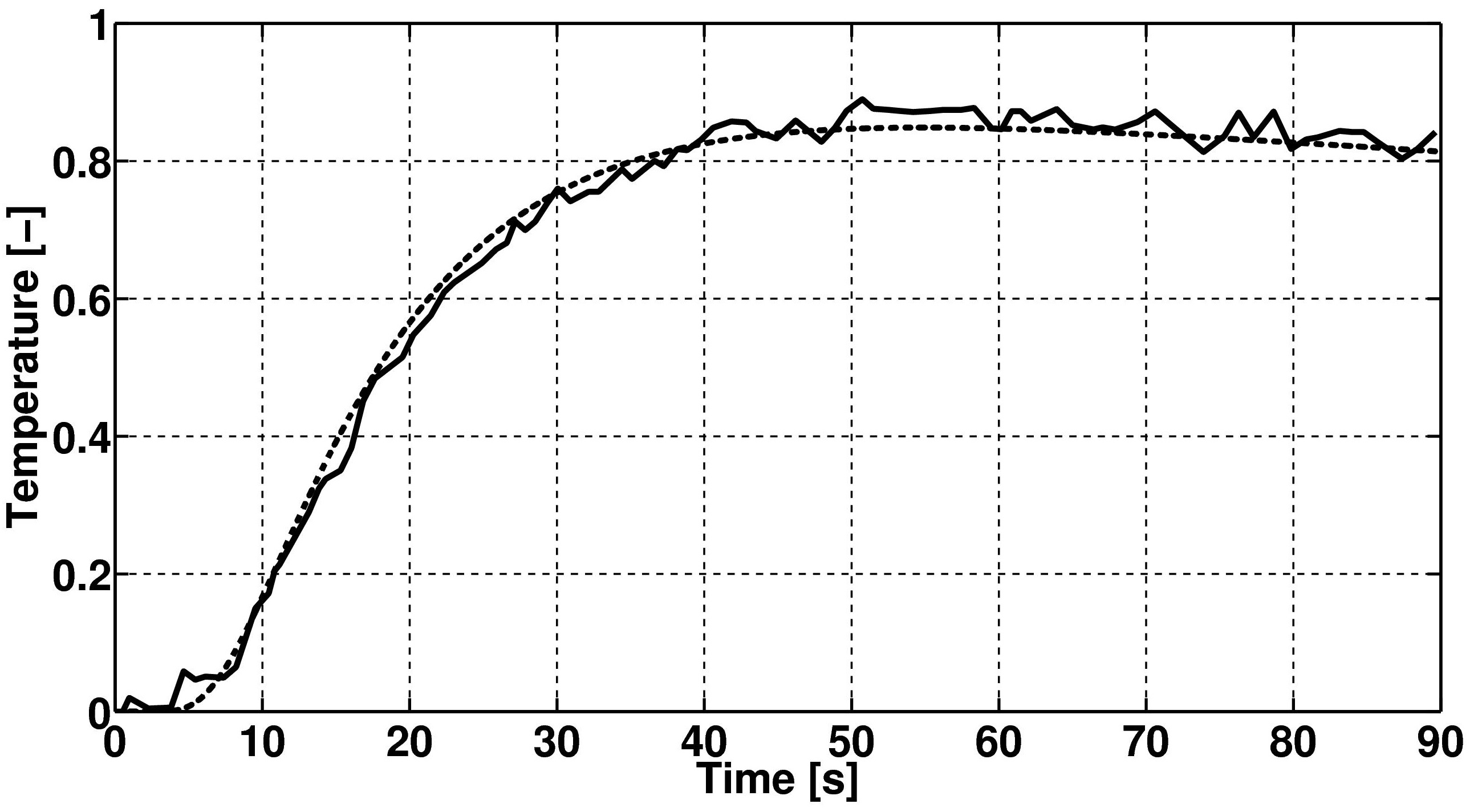}
        }%
        \subfigure[Fourier fit for the third case; $\alpha = 7 \cdot 10^{-8} \f{m^2}{s}$]{%
           \label{fig:tan4}
           \includegraphics[width=6cm,height=3.5cm]{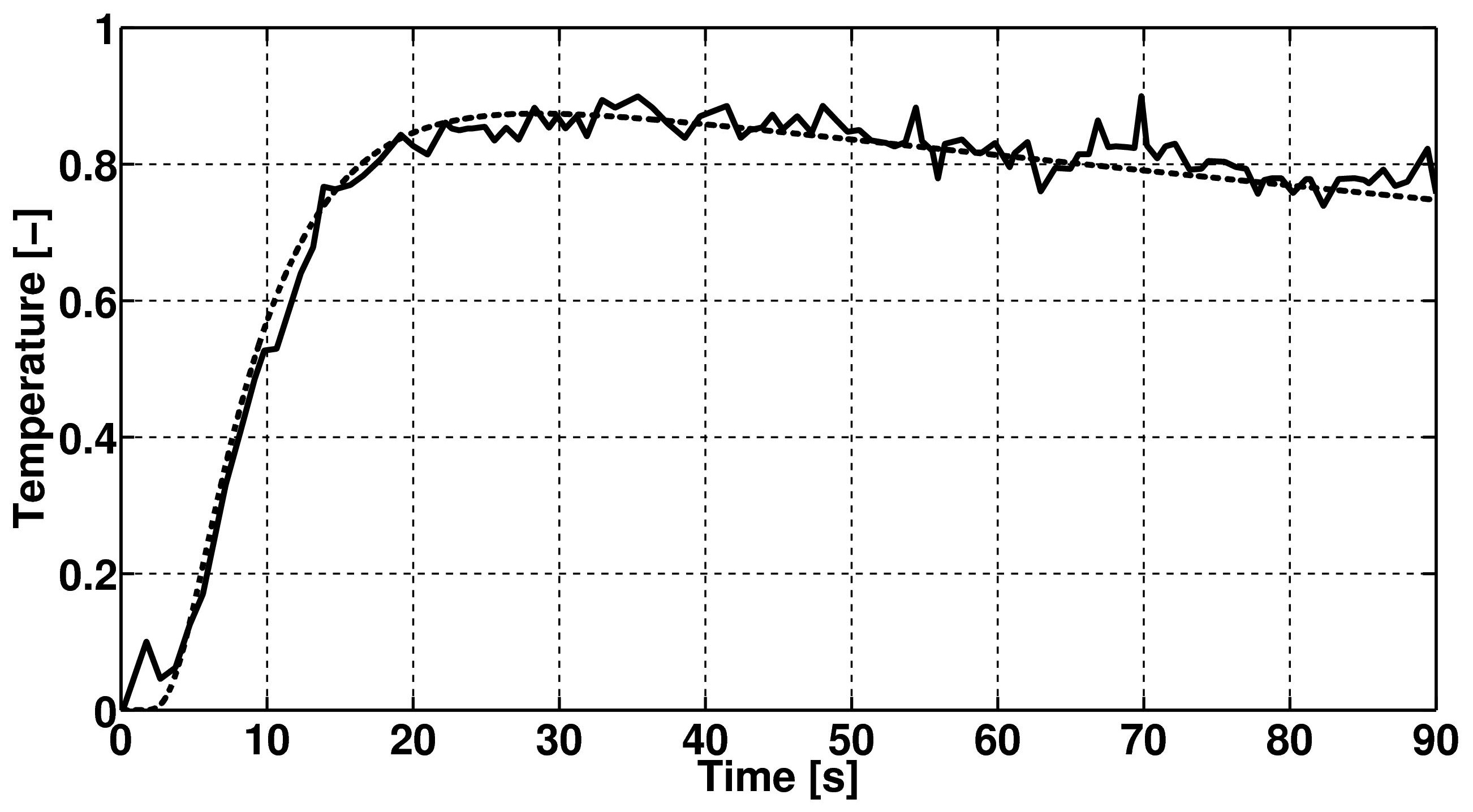}
        }
    \end{center}
    \caption{%
      Solid line denotes the measured data from Tang et al. \cite{TanEtal07}, the dashed line presents the fitting
     }%
   \label{fig:bio_tanetal}
\end{figure}

The fitting is reproduced in the framework of non-equilibrium thermodynamics, see Fig. \ref{fig:bio_tanetal}. It is clear that in the first case the deviation from Fourier's law is real. However, when cooling is present, the Fourier fitting of Tang et al. \cite{TanEtal07} is not appropriate and there is no such deviation in the experimental data, the Fourier's law is applicable in these cases, see Figs. \ref{fig:tan3} and \ref{fig:tan4}. Moreover, as it can be seen on Fig. \ref{fig:bio8}, the deviation is considered as a wave propagation using the DPL model which produces an overshoot instead of a cooling tail \cite{TanEtal07} and it leads to equilibrium below the dimensionless temperature one.

\section{Summary}
\label{SUM}

Heat conduction models with phase lags are presented, compared and their thermodynamical background is discussed. It is concluded that the differential DPL model can be interpreted in the framework of non-equilibrium thermodynamics and this interpretation removes the most important paradoxes. 
\begin{itemize}
\item The GK, the JT and the experimentally important differential DPL equations can be derived in a uniform background. 
\item  The conditions of the second law lead to non-negative coefficients in the equation, independently of obscure causality arguments. 
\item  These pure thermodynamic conditions ensure the expected asymptotic stability of the homogeneous equilibrium without any further conditions. 
\item  The time shift paradox of the DPL does not apply, because the completely different interpretation. 
\item  The validity of the GK equation is extended beyond the phonon interpretation.
\end{itemize}

Further theoretical analysis and  experiments are necessary  to understand the difference of the GK and JT/DPL phenomena and their possible common modeling power. The difference in experimental observation is expected in more than a single spatial dimensions. The theoretical analysis can be informative with the solution of the equations and also in the difference in the universal theoretical interpretations \cite{VanFul12,KovVan15,PMMar17conf}.


\bibliographystyle{spphys}       

%
%


\end{document}